\documentclass[aps,prl,twocolumn,superscriptaddress]{revtex4}
\usepackage{graphicx,amssymb}
\usepackage{bm}
\newcommand\IPA{IPA-CuCl$_3$~}

\begin{document}

\title{Dynamics of composite Haldane spin chains in \IPA}

\author{T. Masuda}
\email[]{tmasuda@yokohama-cu.ac.jp}
\altaffiliation{Present
address: International Graduate School of Arts and Sciences,
Yokohama City University, 22-2, Seto, Kanazawa-ku, Yokohama city,
Kanagawa, 236-0027, Japan} \affiliation{Condensed Matter Science
Division, Oak Ridge National Laboratory, Oak Ridge, TN 37831-6393,
USA}

\author{A. Zheludev}
\email[]{zheludevai@ornl.gov}
\affiliation{Condensed Matter
Science Division, Oak Ridge National Laboratory, Oak Ridge, TN
37831-6393, USA}

\author{H. Manaka}
\affiliation{Graduate School of Science and Engineering, 
Kagoshima University, Korimoto, 
Kagoshima 890-0065, Japan}

\author{L.-P. Regnault}
\affiliation{CEA-Grenoble, DRFMC-SPSMS-MDN, 17 rue des Martyrs,
38054 Grenoble Cedex 9, France.}

\author{J.-H. Chung}
\affiliation{NCNR, National Institute of Standards and Technology,
Gaithersburg, Maryland 20899, USA}

\author{Y. Qiu}
\affiliation{NCNR, National Institute of Standards and Technology,
Gaithersburg, Maryland 20899, USA}

\date{\today}

\begin{abstract}
Magnetic excitations in the quasi-one-dimensional antiferromagnet
\IPA\ are studied by cold neutron inelastic scattering. Strongly
dispersive gap excitations are observed. Contrary to previously
proposed models, the system is best described as an asymmetric
quantum spin ladder. The observed spectrum is interpreted in terms
of ``composite'' Haldane spin chains. The key difference from
actual $S=1$ chains is a sharp cutoff of the single-magnon
spectrum at a certain critical wave vector.
\end{abstract}

\pacs{75.10.Jm, 75.25.+z, 75.50.Ee}

\maketitle

Antiferromagnetic 2-leg spin ($S=1/2$) ladders and the
closely related $S=1$ Haldane spin chains \cite{Haldane83, White96}
are an example of quantum disorder and mass generation in extended
spin networks. In these {\it quantum spin liquids},
the magnetism is suppressed due to {\it collective}
zero-point fluctuations and the unique topology of one dimension
(1D),
rather than to finite system size. Spin
ladders play a key role in the dynamics of stripe phases in
high-temperature superconductors~\cite{Tranquada04}
and under certain
conditions can themselves support exotic types of
superconductivity~\cite{Uehara96}.
On a more fundamental level they are ideal models for studying the
collective spin dynamics in 1D, quantum critical points and phase
transitions in external magnetic fields, and the effects of
quenched disorder.

Despite the wealth of relevant theoretical results, experimental
studies are lagging behind for shortage of suitable model
compounds. In the best known ladder examples
Sr$_{14}$Cu$_{24}$O$_{41}$~\cite{Matsuda96} and
SrCu$_2$O$_3$~\cite{Azuma94} the large energy scales of magnetic
interactions limit spectroscopic studies, especially at high
fields. A typical problem with many known Haldane-gap systems
\cite{Regnault94,Zheludev01} is a large single-ion magnetic
anisotropy that is often associated with $S=1$ spins, and
qualitatively affects the dynamics and field behavior
\cite{ZheludevNDMAP}. Moreover, a Haldane spin chain intrinsically
has a smaller Hilbert space than $S=1/2$ spin ladders, and
therefore lacks certain very interesting spectral features
\cite{Sushkov98}. In the present work we report the discovery of
an isotropic ladder spin network in the $S=1/2$ compound
(CH$_3$)$_2$CHNH$_3$CuCl$_3$ (\IPA), that was previously thought to
be a prototypical ferromagnetic-antiferromagnetic (F-AF) spin
chain \cite{Manaka97a}. We use inelastic neutron scattering to
study its magnetic excitation spectrum. The experiments reveal a
spectacular {\it truncation of the magnon branch} at certain
critical wave vectors that we attributed to peculiarities of
magnon interactions, and bring new insight on diluted
(CH$_3$)$_2$CHNH$_3$Cu(Cl$_x$Br$_{1-x}$)$_3$ compounds
\cite{Manaka02}.

\begin{figure}
\begin{center}
\includegraphics[width=8cm]{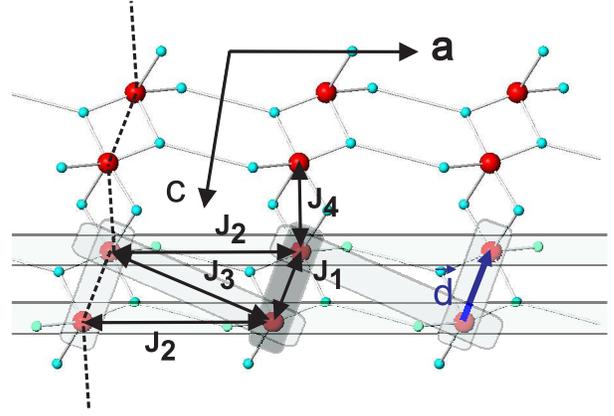}
\end{center}
\caption{Layers of magnetic Cu$^{2+}$ (red) and the bridging
Cl$^-$ ions (cyan) in \IPA. Previously proposed alternating Cu
chains (dotted line) run along the $c$ axis. Actual asymmetric
spin ladders (shaded) are parallel to the $a$ direction.
The
ladder rungs are defined by the vector $\mathbf{d}$.
}
\label{fig1}
\end{figure}

\IPA crystallizes in a triclinic space group $P\bar{1}$ with $a =
7.766$ \AA , $b = 9.705$ \AA , $c = 6.083$ \AA , $\alpha =
97.62^{\circ}$, $\beta = 101.05^{\circ}$, and $\gamma =
67.28^{\circ}$~\cite{Manaka97a}. The key features of the structure are shown in
Fig.~\ref{fig1}. The magnetism is due to $S=1/2$-carrying
Cu$^{2+}$ ions arranged in sheets parallel to the $(a,c)$
crystallographic plane. These sheets are well separated by
non-magnetic organic layers. A spin gap of $\Delta\approx 1.5$~meV
in \IPA\ was first discovered by Manaka {\it et
al.}~\cite{Manaka97a} in $\chi(T)$ measurements. This
value is consistent with the critical field $H_\mathrm{c}\sim
11$~T that induces an ordered antiferromagnetic phase at low
temperatures~\cite{Manaka98a}. The gap was attributed to a singlet
ground state of bond-alternating Cu$^{2+}$ chains running along
the crystallographic $c$ axis, as shown by dotted line in
Fig.~\ref{fig1}. Indeed, this model consistently explained all
magnetization curves and ESR experiments, with a quantitative
agreement obtained assuming alternating F-AF
bonds~\cite{Manaka97a}. An F coupling between
nearest-neighbor Cu$^{2+}$ sites 
$J_1 < 0$~\cite{footnote} 
is consistent with the relevant
Cu-Cl-Cu bond angles in the crystal structure of \IPA.
\begin{figure*}
\begin{center}
\includegraphics{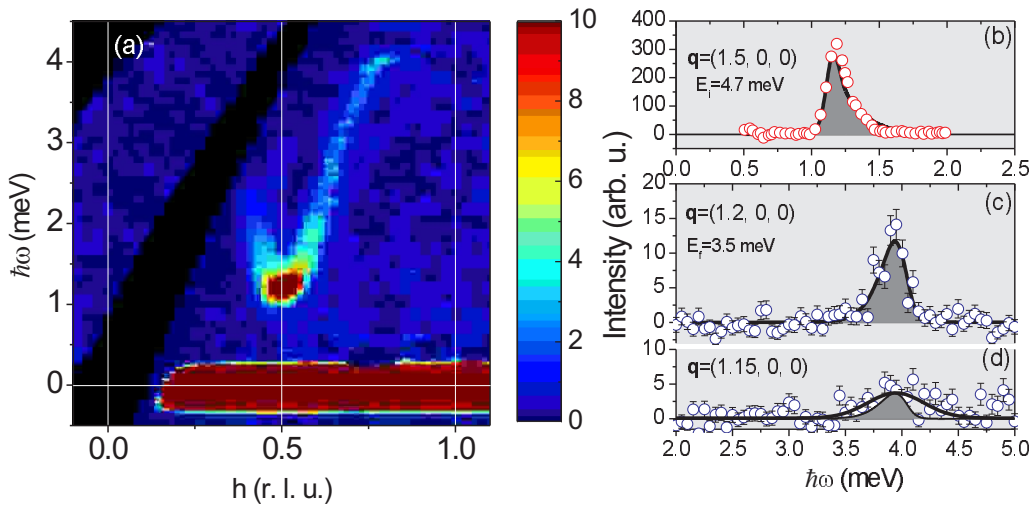}
\end{center}
\caption{Time of flight (a) and 3-axis (b)-(d) inelastic neutron
data measured in \IPA at $T = 1.5$ K. In (a) the suppression of inelastic
intensity at $h\le 0.5$ in (a) is due to a large $l$ (momentum
transfer along $c^\ast$) and the FM-dimer structure factor for the
magnon branch (see text). For $h>0.5$, $l$ is small at all times :
$|l|<0.2$. In (b) - (d) the shaded areas are calculated peak
shapes due to resolution. } \label{fig2}
\end{figure*}

At the center of the present work are inelastic neutron scattering
experiments on deuterated \IPA\ samples, prepared by
crystallization from solution~\cite{Manaka97a}. 
The $\chi (T)$ and ESR data for the deuterated
samples agree well with those for the non-deuterated ones. 
20 single crystals
with total mass of 3.5g were co-aligned to a mosaic spread of
~4$^\circ$. The data were collected using cold-neutron 3-axis
spectrometers NG5-SPINS in NIST and IN12 in ILL, and at the Disc
Chopper time-of-flight Spectrometer (DCS) at NIST~\cite{Copley}. The sample was
mounted with the $(a,c)$ crystallographic plane parallel to the
scattering plane of the instruments and maintained at or below
$T=1.5$~K. On SPINS we utilized
$(\mathrm{guide})-80'-80'-(\mathrm{open})$ collimations and a BeO
filter after the sample for a fixed final neutron energy
$E_\mathrm{f}=3.7$~meV, selected by a flat pyrolitic graphite (PG)
analyzer. On IN-12 a Soller collimator of $60'$ was used only at
pre-sample position. A Be filter was positioned after the sample.
Neutrons of fixed-incident $E_\mathrm{i}$ = 4.7 meV or fixed-final
$E_\mathrm{f}$ = 3.5 meV energies were used in conjunction with  a
horizontally focusing PG analyzer. A constant background
(typically 1.5 counts/min, depending on configuration) was
subtracted from all 3-axis data sets. On the DCS instrument the
data were taken with $E_\mathrm{i}= 6.7$~meV neutrons, the incident
beam forming an angle of $60^\circ$ with the $a^\ast$ direction.
The background was directly measured by removing the sample from
the cryostat.

\begin{figure}
\begin{center}
\includegraphics[width=8.7cm]{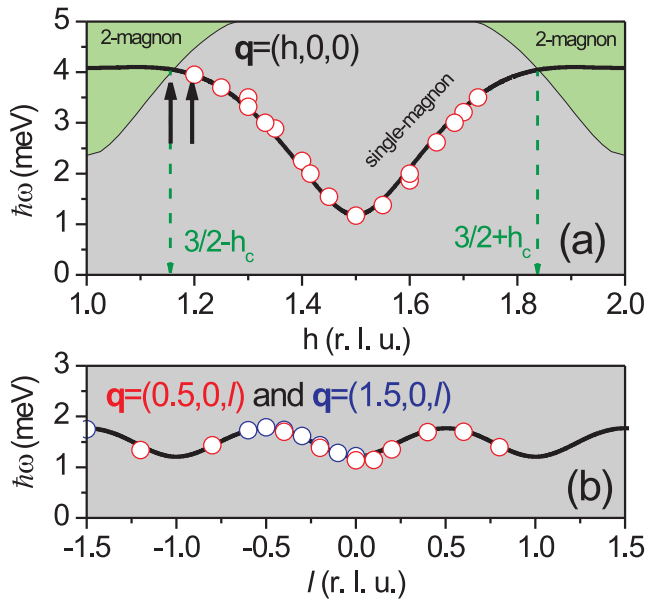}
\end{center}
\caption{Magnon dispersion measured in \IPA along the $a^\ast$ (a)
and $c^\ast$ (b) directions (symbols). Heavy solid lines are fits, as
described in the text. In (a) the green shaded areas indicate the
domain of the 2-magnon continuum. Solid arrows show the positions
of scans in Fig.~\protect\ref{fig2} (c) and (d). Dashed arrows show the
critical wave vectors.} \label{fig3}
\end{figure}

A series of constant-$E$ and constant-$q$ scans revealed
well-defined long-lived magnetic gap excitations in a large part
of the Brillouin zone in IPA-CuCl$_3$. Typical data are shown in
Fig.~\ref{fig2}. Our measurements unambiguously show that the
magnetic strong-coupling direction in \IPA~ is along the crystallographic
$a$ axis. The global dispersion minimum is at
$\mathbf{q}=(\case{2n+1}{2},0,0)$ with $n$-integer (Fig.~\ref{fig3}a),
and also corresponds to a maximum of observed inelastic intensity.
In contrast, the dispersion along the $c^*$ axis is rather weak,
with an energy minimum at $l=0$ in Fig.~\ref{fig3}b. The
originally proposed model of bond-alternating F-AF chains running
along the crystallographic $c$ axis is thus {\it totally
inconsistent} with our results. Nevertheless, the measured gap
energy $\Delta \sim 1.2$~meV is in agreement with that  deduced
from bulk measurements.

\begin{figure}
\begin{center}
\includegraphics[width=8.7cm]{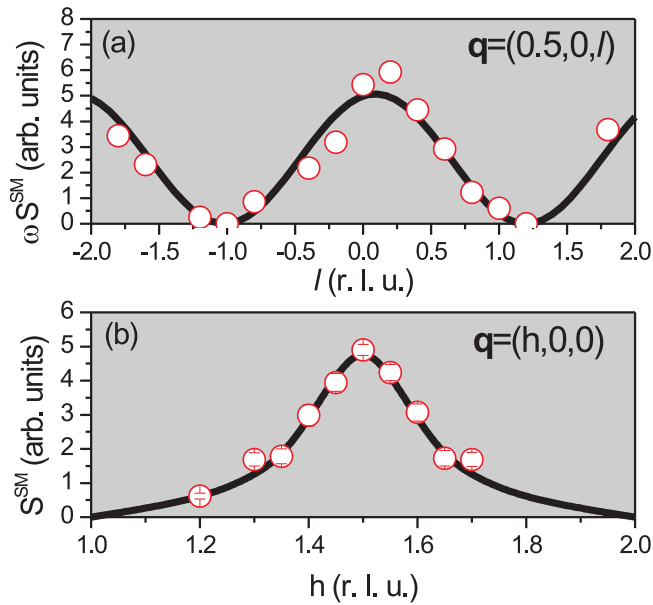}
\end{center}
\caption{Measured $l$- (a) and $h$- dependencies (b) of
single-magnon excitation intensities in \IPA (symbols). Solid
lines are simulations, as described in the text.} \label{fig4}
\end{figure}

In the wave vector range $0.2 \le h \le 0.8$ and $1.2 \le h \le
1.8$, where sharp magnon peaks were observed, the 3-axis data
were analyzed using a model single-mode (SM) cross section, with an 
empirical spin-gap dispersion relation~\cite{Barnes}, 
slightly modified to accomodate a transverse disperion:
\begin{eqnarray}
 S(\mathbf{q},\omega)&\propto& S^{\mathrm{SM}}(\mathbf{q})
 \delta(\omega-\omega_\mathbf{q}),\\
 \omega_\mathbf{q}^2&=&\omega_0^2\cos^2(\pi h) +  \left[\Delta^2+4b_0^2\sin^2(\pi l)\right]\sin^2(\pi
 h) + \nonumber\\
 &+& c_0^2\sin^2(2\pi h),\label{disp}
\end{eqnarray}
where
$\mathbf{q}=h\mathbf{a}^\ast+k\mathbf{b}^\ast+l\mathbf{c}^\ast$.
The gap is equal to $\Delta$ at $l=0$, and is allowed to slightly
oscillate along $c^\ast$ with a transverse bandwidth $b_0$. 
The boundary energy corresponds to $\omega _0$. 
The cross section was numerically convoluted with the spectrometer
resolution function calculated in the Popovici
approximation\cite{Popovici} and fit to the experimental data. The
magnetic form factor for Cu$^{2+}$ ions was built into the fits.
Separate coefficients $S^{\mathrm{SM}}(\mathbf{q})$ were used for
each constant-$\mathbf{q}$ scan to determine the 3-dimensional
structure factor in the most model-independent manner possible.
The parameters of the dispersion relation in Eq.~(\ref{disp}) were
treated as global for all the collected scans. The fitting
procedure yielded  $\omega_0 = 4.08$ (9) meV, $\Delta = 1.17$
(1) meV, $b_0=0.67$ (1)~meV and $c_0 = 2.15$ (9) meV. Scans
simulated using these parameter values are shown as shaded areas
in Figs.~\ref{fig2}b-d. The dispersion relations are plotted in
solid lines in Fig.~\ref{fig3}. Symbols in these plots were
obtained in fits to individual scans, as opposed to global fits to
the measured data. Figure~\ref{fig4}a shows the $l$-dependence of
the single-mode component of the dynamic structure factor
$S^{\mathrm{SM}}(\mathbf{q})$ measured at $h=0.5$.  The
$h$-dependence of this structure factor for $l=0$ is shown in
Fig.~\ref{fig4}b.

The measured $c$-axis modulation of the magnon intensity is
exactly explained by a ferromagnetic dimer with a particular spin
separation $\mathbf{d}$ in the crystal structure
(Fig.~\ref{fig1}): $\mathbf{d} = (-0.0854,~-0.1316,~0.4432)$. The
data on the $h=\case{2n+1}{2}$ reciprocal-space rods are very well
reproduced by assuming $S^{\mathrm{SM}}(\mathbf{q})\propto
\cos^2(\pi hd_x + \pi ld_z)/\omega_\mathbf{q}$, as shown by the
solid line in Fig.~\ref{fig4}a. To a good approximation one can
thus describe \IPA in terms of uniform chains made up of these
ferromagnetically correlated spin pairs. The chains run along the
crystallographic $a$ axis and are composed of effective $S=1$
spins. Since the single mode intensity has its maximum at
$h=\case{2n+1}{2}$, (Fig.~\ref{fig4}b), the correlations between
the nearest-neighbor effective $S=1$ spins are predominantly
antiferromagnetic. If one assumes that near the magnetic
zone-center a dominant fraction of the total spectral weight is
contained in the single mode excitations, one can reproduce the
measured $h$-dependence of intensity with only nearest-neighbor AF
interactions in the $S=1$ chains. For this model $S(h)\propto
\sin^{2}(\pi h) / \omega_h$, which is in good agreement with
experiment (solid line in Fig.~\ref{fig4}b). We conclude that the
singlet ground state in \IPA and the gap excitations are due to
{\it composite $S=1$ Haldane spin chains} that run {\it
perpendicular} to the originally proposed bond-alternating chains
directions.

As shown in Fig.~\ref{fig1} the crystal structure of \IPA presents
a number of possible AF superexchgange routes between Cu$^{2+}$
spins along the $a$ axis. In the most general case, the
appropriate model for \IPA is a {\it spin ladder} with F rung
interactions $J_1$, a leg coupling constant $J_2$ and a diagonal
exchange interaction of magnitude $J_3$ (presumably also AF). 
If this interaction was infinitely strong, the mapping
onto a Haldane spin chain would be exact. In fact, the measured
dispersion differs from that for an ideal $S=1$ chain. In the
latter, the ratio of the spin wave velocity $v$ to the gap energy
$\Delta$ is $v/\Delta \sim 6$ \cite{Sorensen94}. In contrast, in \IPA
$v/\Delta \approx 3.1$, about twice as small. The analogy with
Haldane spin chains is thus only qualitative, and the $|J_1|$ is
probably of the same order of magnitude as $J_2$ and $J_3$.
Without a direct comparison with first-principle calculations, the
exact values of the exchange constants can not be deduced from the
measured dispersion and intensity modulation of the single-magnon
excitations. The observed small dispersion of magnetic excitations
along the $c^\ast$ direction indicates a weak coupling $J_4$
between individual ladders (Fig.~\ref{fig1}). From the position of
the dispersion minimum we conclude that 
$J_4<0$.

A remarkable feature of the measured spectrum is the abrupt
disappearance of the magnon branch at a certain critical wave
vector $q_\mathrm{c}$. This phenomenon is well illustrated by the
time of flight data of Fig.~\ref{fig2}a. The sharp gap excitation
extends to $h=\frac{1}{2}+h_\mathrm{c} \sim 0.8$, where $h_\mathrm{c}
\sim 0.3$, and then vanishes abruptly, rather than persisting to
the zone-boundary at $h=1$. Constant-$q$ 3-axis data reveal
similar behavior at the equivalent wave vector
$h=\frac{3}{2}-h_\mathrm{c}\sim 1.2$. The well-defined resolution
limited peak seen at $h=1.2$ (Fig.~\ref{fig2}c) is weakened and
broadened at $h=1.15$ (Fig.~\ref{fig2}d, where the shaded area is
the simulated profile for a resolution-limited peak), and is
totally absent at $h=1.1$ (not shown).

Such a dramatic {\it spectrum termination} phenomenon has been
recently discovered and investigated in the quasi-2D material PHCC
\cite{Stone2005}. To our knowledge, \IPA is the only other known
spin gap material with a similar feature. The effect can be
explained by a coupling between single-magnon and multi-magnon
states \cite{Stone2005,Zhitomirsky}. The domain of the two-magnon
continuum in \IPA, though invisible in our experiments due to low
intensity and a very tight energy resolution, can be calculated
using the measured one-magnon dispersion. As represented by the
green shaded areas in Fig.~\ref{fig3}a, the continuum has a gap of
$2\Delta$. Note that the single-magnon branch crosses into the
2-magnon continuum at exactly $h= \frac{2n+1}{2}\pm h_\mathrm{c}$.
Energy and momentum conservation laws forbid any single-magnon
excitation {\it outside} the domain to decay into bunches of
multiple magnons. For $\frac{2n+1}{2}- h_\mathrm{c} < h
<\frac{2n+1}{2}+ h_\mathrm{c}$ the magnons in \IPA are thus
stable. Their contribution to the dynamic structure factor are
sharp $\delta$-function peaks. This restriction is removed for
magnons inside the continuum: they are rendered unstable through
decaying into pairs of lower-energy magnons. Their energy width
(inverse lifetime) is dramatically increased. In \IPA this effect
is particularly strong, and for $\frac{2n-1}{2}+ h_\mathrm{c} < h
<\frac{2n+1}{2}- h_\mathrm{c}$ there are no detectable
single-magnon states.

What drives the decay process is interactions between magnons.
Interestingly, in an ideal symmetric spin ladder the matrix
elements between single-magnon and 2-magnon states are exactly
zero, so the single mode survives inside the 2-magnon continuum
\cite{Zhitomirsky}.
In particular, in an isotropic Haldane spin chain stable magnons
are expected to persist all the way to the zone-boundary. We can
speculate that the abrupt disappearance of the magnon branch in
\IPA, which distinguishes these composite Haldane spin chains from
actual ones,  is due to a symmetry-breaking diagonal interaction
$J_3$.

We finally want to point out the relevance of our results  to the
behavior of isostructural mixed
(CH$_3$)$_2$CHNH$_3$Cu(Cl$_x$Br$_{1-x}$)$_3$ materials
\cite{Manaka02}. In these compounds the spin gap decreases with
increasing $x$, closes at $x\approx 0.44$ (the system becomes
magnetically ordered), and then re-opens at $x>0.87$. It was
proposed that Br doping can change $J_1$ from $J_1<0$ at $x=1$ to
$J_1>0$ at $x=0$ \cite{Manaka02}. The phase diagram of such
bond-disordered ladders has not been studied to date. However, on
a naive mean-field level we can think of doping as a way to
continuously tune $J_1$. In this case, increasing $x$ should lead
the system through a gapless quantum-critical state, which for
$J_3=0$ is exactly at $J_1=0$. The reenterant doping phase diagram
is then easy to understand. Note that in the previously proposed
alternating-chain model the gap remains open at $J_1=0$.

Contrary to what was previously thought, \IPA is a beautiful
asymmetric spin ladder system, that for many purposes can be seen
as a ``composite'' Haldane spin chain. Unlike an ideal Haldane
spin chain though, it features a spectacular end-of-spectrum
effect at a certain critical wave vector. Future neutron work will
focus on a direct observation of excitation continua, effects of
high magnetic field and Br-dilution.

We thank M. Stone, I. Zaliznyak and M. Zhitomirsky for insightful
discussions. 
S. Raymond is appreciated for his experimental support at ILL. 
Work at ORNL was carried out under Contracts No.
DE-AC05-00OR22725, US Department of Energy. T. M. was partially
supported by the US - Japan Cooperative Research Program on
Neutron Scattering between the US DOE and Japanese MEXT.
The work at NIST is supported by 
the National Science Foundation under Agreement Nos. DMR-9986442, 
-0086210, and -0454672. 




\begin{thebibliography}{20}
\bibitem{footnote}
In our definition $J>0$ corresponds to AF interactions.
The 2-spin Hamiltonian is written as $+J\mathbf{S}_1\mathbf{S}_2$.
Note that Refs.~\protect\cite{Manaka97a,Manaka02,Manaka98a} uses a
different convention: $J>0$ corresponds to ferromagnetism and the
exchange term is $-J\mathbf{S}_1\mathbf{S}_2$.
\end{thebibliography}

\end{document}